\documentclass[aps,prb,10pt,twocolumn,superscriptaddress,amsmath,amssymb]{revtex4-2}

\usepackage{graphicx} 
\usepackage{dcolumn} 

\def\bk{{\bf k}}

\def\bq{{\bf q}}

\def\bQ{{\bf Q}}

\def\b0{{\bf 0}}

\def\cO{{\cal O}}

\def\Re{{\rm Re}}
\def\Im{{\rm Im}}

\def\alf{\alpha}
\def\eps{\epsilon}
\def\gam{\gamma}
\def\Gam{\Gamma}
\def\lam{\lambda}
\def\Lam{\Lambda}
\def\om{\omega}

\def\sg{\sigma}
\def\Sg{\Sigma}

\def\tk{\tilde k}
\def\tq{\tilde q}

\def\tom{\tilde\om}

\begin{document}

\title{Marginal Fermi liquid behavior at the onset of $\boldsymbol{2k_F}$ density wave order \\ in two-dimensional metals with flat hot spots}

\author{Lukas Debbeler}
\affiliation{Max Planck Institute for Solid State Research,
 D-70569 Stuttgart, Germany}
\author{Walter Metzner}
\affiliation{Max Planck Institute for Solid State Research,
 D-70569 Stuttgart, Germany}

\date{\today}

\begin{abstract}
We analyze quantum fluctuation effects at the onset of incommensurate $2k_F$ charge- or spin-density wave order in two-dimensional metals, for a model where the ordering wave vector $\bQ$ connects a single pair of hot spots on the Fermi surface with a vanishing Fermi surface curvature. The tangential momentum dependence of the bare dispersion near the hot spots is proportional to $|k_t|^\alf$ with $\alf > 2$. We first compute the order parameter susceptibility and the fermion self-energy in random phase approximation (RPA). Logarithmic divergences are subsequently treated by a renormalization group analysis. The coupling between the order parameter fluctuations and the fermions vanishes logarithmically in the low-energy limit. As a consequence, the logarithmic divergences found in RPA do not sum up to anomalous power laws. Instead, only logarithmic corrections to Fermi liquid behavior are obtained. In particular, the quasiparticle weight and the Fermi velocity vanish logarithmically at the hot spots.
\end{abstract}

\maketitle


\section{Introduction} \label{sec:Introduction}

Quantum fluctuations at and near quantum critical points in metallic electron systems can trigger non-Fermi liquid behavior with unconventional temperature, momentum, and frequency dependencies of thermodynamic, spectroscopic, and transport properties \cite{loehneysen07}. In view of non-Fermi liquid or ``strange metal'' behavior observed in several layered compounds such as the high-$T_c$ cuprates, quantum criticality in two-dimensional systems has attracted particular interest.

Metals at the onset of charge or spin-density wave order can be grouped in several distinct universality classes of quantum critical non-Fermi liquids. The most thoroughly studied case of N\'eel order is just one example \cite{abanov03,metlitski10_af1,lee18,berg12}.
A particularly intriguing situation arises when the wave vector $\bQ$ of the density wave is a {\em nesting vector}\/ (also known as ``$2k_F$'' vector \cite{altshuler95}) of the Fermi surface, that is, when it connects Fermi points with collinear Fermi velocities \cite{footnote_perfnest}. Charge and spin susceptibilities exhibit a singularity at such wave vectors due to an enhanced phase space for low-energy particle-hole excitations.
The wave vector of a N\'eel state is a nesting vector only for special electron densities \cite{bergeron12,wang13}.
While fluctuations are naturally stronger in two dimensions, quantum fluctuation effects at the onset of $2k_F$ density wave order are interesting also in three dimensions \cite{schaefer17}.

Non-Fermi liquid behavior at the onset of charge- or spin-density wave order with {\em incommensurate}\/ \cite{fn_incomm} nesting wave vectors $\bQ$ in two-dimensional metals has already been analyzed, too.
In a perturbative one-loop calculation of the fermion self-energy, a breakdown of Fermi liquid behavior was found at the {\em hot spots}\/ on the Fermi surface connected by the ordering wave vector \cite{holder14}. If the ordering wave vector $\bQ$ connects only a single pair of hot spots, in axial or diagonal direction, the frequency dependence of the one-loop self-energy at the hot spots obeys a power-law with exponent~$\frac{2}{3}$. If $\bQ$ connects two pairs of hot spots, the imaginary part of the real frequency one-loop self-energy exhibits a linear frequency dependence.
In none of these two cases the perturbative solution is self-consistent, and the feedback of the non-Fermi liquid self-energy seems to shift the ordering wave vector away from the nesting point \cite{sykora18,sykora21}. For the case of a single hot spot pair, it was argued already long ago that quantum fluctuations replace the QCP by a first order transition \cite{altshuler95}.
However, a fluctuation induced flattening of the Fermi surface at the hot spots might save the QCP \cite{sykora18}, and this scenario was supported by a systematic $\epsilon$-expansion around the critical dimension $d_c = \frac{5}{2}$ \cite{halbinger19}.
For the two hot-spot pair case, a self-consistent solution with a stable QCP was found numerically \cite{sykora21}.

Recently, we have analyzed non-Fermi liquid behavior at the onset of density wave order for a case where the nesting vector connects a single pair of {\em flat}\/ hot spots, where the Fermi surface curvature vanishes already in the non-interacting reference system, that is, before fluctuations are taken into account \cite{debbeler23}. Such a situation can arise at special electron filling factors, for example, in a tight-binding model with nearest and next-nearest neighbor hopping on a square lattice.
\begin{figure}[tb]
\centering
\includegraphics[width=7cm]{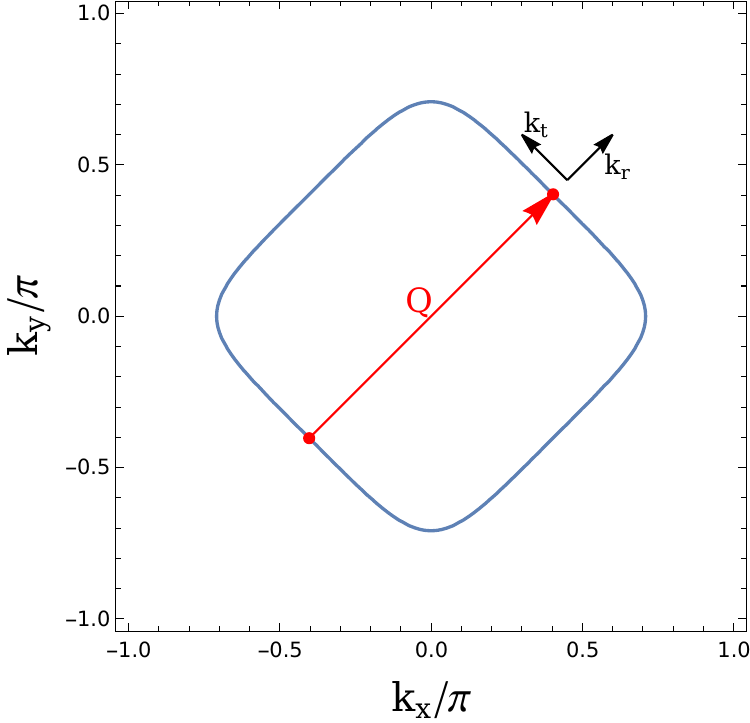}
\caption{Hot spots with vanishing Fermi surface curvature connected by the ordering wave vector $\bQ$, and the relative momentum coordinates $k_r$ and $k_t$.}
\label{fig:hotspots}
\end{figure}
Expanded in terms of relative momentum coordinates $k_r$ and $k_t$ in normal and tangential directions with respect to the Fermi surface at a hot spot (see Fig.~\ref{fig:hotspots}), the dispersion relation assumed in Ref.~\cite{debbeler23} has the form $\xi_\bk = \eps_\bk - \mu = v k_r + b k_t^4$, where $v$ is the Fermi velocity at the hot spot, and $b$ is a real constant.
Unlike the conventional case of hot spots with finite curvature, the order parameter susceptibility of our model with flat hot spots has a robust peak at the nesting vector. The imaginary part of the perturbative one-loop self-energy was found to depend linearly on (real) frequency, while the momentum and frequency dependences of its real part exhibit logarithmic divergences indicating non-Fermi liquid behavior with a vanishing quasiparticle weight and a vanishing Fermi velocity. The logarithmic divergences were tentatively interpreted as perturbative signatures of power-laws with anomalous exponents \cite{debbeler23}.

In this paper we extend our previous analysis of non-Fermi liquid behavior at $2k_F$ density wave quantum criticality with flat hot spots in two directions. First, we treat the logarithmic divergences obtained in perturbation theory in a controlled and systematic way by using a field theoretic renormalization group. Second, we generalize the tangential momentum dependence of the bare dispersion from quartic to arbitrary powers, that is,
\begin{equation} \label{dispersion}
 \xi_\bk = \eps_\bk - \mu = v k_r + b|k_t|^\alf \, ,
\end{equation}
with an arbitrary exponent $\alf > 2$. Our motivation for this generalization is to disentangle general features of models with flat hot spots from features specific for the case $\alf = 4$. Moreover, we would like to explore whether one can shed light on the most important but also most tricky case $\alf = 2$ by considering the limit $\alf \to 2$.

The renormalization group analysis reveals that the logarithms found in perturbation theory do not sum up to anomalous power-laws, but rather yield only a logarithmic breakdown of Fermi liquid theory. Hence, the flat hot spot model turns out to be a {\em marginal}\/ Fermi liquid \cite{varma89}.
For $\alf > 2$ the renormalization group analysis is controlled by the inverse number of fermion flavors $N$. However, in the limit $\alf \to 2$, the fluctuation corrections exhibit divergences which cannot be tamed by the renormalization group, indicating a qualitatively different behavior for $\alf = 2$.

Recently, Song \cite{song23} et al.\ have computed the quasiparticle decay rate near Fermi points with a dispersion of the form Eq.~\eqref{dispersion} in a stable two-dimensional Fermi liquid (away from instabilities), and found an energy dependence of the form $\eps^{\frac{\alf}{\alf-1}}$ for $\alf > 2$. While this decay rate is enhanced compared to the conventional quadratic behavior, it is still parametrically smaller than the quasiparticle energy in the low energy limit, for any finite $\alf$, so that the quasiparticles remain asymptotically stable.

The remainder of the paper is structured as follows. In Sec.~II we compute the order parameter susceptibility and the effective interaction at the QCP in one-loop approximation. A perturbative one-loop calculation of the momentum and frequency dependence of the fermion self-energy is performed in Sec.~III, and the corresponding renormalization group analysis in Sec.~IV. In Sec.~V we close the presentation with a summary and discussion of the main results.


\section{Susceptibility and effective interaction} \label{sec:Susceptibility}

We consider a one-band system of interacting fermions with a bare single-particle energy-momentum relation $\eps_\bk$. We are dealing exclusively with ground state properties (temperature $T=0$). The bare fermion propagator has the form
\begin{equation}
 G_0(\bk,ik_0) = \frac{1}{ik_0 - \xi_\bk} \, ,
\end{equation}
where $k_0$ denotes the imaginary frequency, and $\xi_\bk = \eps_\bk - \mu$.
We assume that, in mean-field theory, the system undergoes a charge or spin-density wave transition with an incommensurate and nested wave vector $\bQ$, which connects a pair of hot spots on the Fermi surface, where the dispersion relation in the vicinity of the hot spots has a momentum dependence of the from Eq.~\eqref{dispersion}.

In random phase approximation (RPA) the order parameter susceptibility has the form
\begin{equation} \label{chi}
 \chi(\bq,iq_0) = \frac{\chi_0(\bq,iq_0)}{1 + g \chi_0(\bq,iq_0)} \, ,
\end{equation}
where $g<0$ is the coupling constant parametrizing the interaction in the instability channel.  The bare charge or spin susceptibility $\chi_0$ is related to the particle-hole bubble $\Pi_0$ by $\chi_0(\bq,iq_0) = - N \Pi_0(\bq,iq_0)$, where $N$ is the spin multiplicity, and \cite{negele87}
\begin{equation} \label{Pi0def}
 \Pi_0(\bq,iq_0) = \int_\bk \int_{k_0} G_0(\bk,ik_0) \, G_0(\bk-\bq,ik_0-iq_0) \, .
\end{equation}
$\int_\bk$ is a shorthand notation for $\int \frac{d^2\bk}{(2\pi)^2}$, and $\int_{k_0}$ for $\int \frac{dk_0}{2\pi}$.
While keeping $N$ as a general parameter in our equations, we choose $N=2$, corresponding to spin-$\frac{1}{2}$ fermions, in all numerical results.
Continuing $\Pi_0(\bq,iq_0)$ analytically to the real frequency axis from the upper complex frequency half-plane yields the retarded polarization function $\Pi_0(\bq,\om)$.

The RPA susceptibility diverges when $g \chi_0(\bQ,0) = -1$, signalling an instability at the critical coupling $g_c = -1/\chi_0(\bQ,0)$ toward charge or spin density wave order with one of the nesting wave vectors $\bQ$ at which $\chi_0(\bq,0)$ has a (finite) peak.

To analyze the behavior of the susceptibility near the singularity, we expand
\begin{equation}
 \delta\Pi_0(\bq,\om) = \Pi_0(\bq,\om) - \Pi_0(\bQ,0) \, .
\end{equation}
for $\bq$ near $\bQ$ and small $\om$. Momenta near $\bQ$ are parametrized by relative momentum coordinates $q_r$ and $q_t$, parallel and perpendicular to $\bQ$, respectively.
The leading contributions to $\delta\Pi_0(\bq,\om)$ come from fermionic momenta near the hot spots connected by $\bQ$, where the dispersion relations in Eq.~\eqref{Pi0def} can be expanded as in Eq.~\eqref{dispersion}, that is, $\xi_\bk = v k_r + b |k_t|^\alf$ and
$\xi_{\bk-\bq} = - v (k_r - q_r) + b |k_t - q_t|^\alf$.
In the following we assume that $b$ is positive. Our derivations and results can be easily adapted to negative $b$.

For $q_t = 0$ all integrations are elementary, and we obtain
$\delta\Pi_0(q_r,0,\om) = \delta\Pi_0^+(q_r,0,\om) + \delta\Pi_0^-(q_r,0,\om)$,
where
\begin{equation}
 \delta\Pi_0^+(q_r,0,\om) =
 \frac{|\om - vq_r|^{\frac{1}{\alf}}}{4\pi v(2b)^{\frac{1}{\alf}}} \times
 \left\{ \begin{array}{ll}
 \cot\frac{\pi}{\alf} - i & \mbox{for} \;\, \om > vq_r \, , \\[3mm]
 (\sin\frac{\pi}{\alf})^{-1} & \mbox{for} \;\, \om < vq_r \, ,
 \end{array} \right.
\end{equation}
and $\delta\Pi_0^-(q_r,0,\om) = [\delta\Pi_0^+(q_r,0,-\om)]^*$.
In the static limit $\om \to 0$, this yields
\begin{equation} \label{dPi0stat}
 \delta\Pi_0(q_r,0,0) =
 \frac{|vq_r|^{\frac{1}{\alf}}}{2\pi v(2b)^{\frac{1}{\alf}}} \times
 \left\{ \begin{array}{ll}
 \cot\frac{\pi}{\alf} & \mbox{for} \;\, q_r < 0 \, ,  \\[3mm]
 (\sin\frac{\pi}{\alf})^{-1} & \mbox{for} \;\, q_r > 0 \, .
 \end{array} \right.
\end{equation}
$\delta\Pi_0(q_r,0,0)$ has a cusp with diverging slope for $q_r \to 0$ for any $\alf > 2$. In the special case $\alf=2$, the slope vanishes for $q_r < 0$.

\begin{figure}[tb]
\centering
\includegraphics[width=8.5cm]{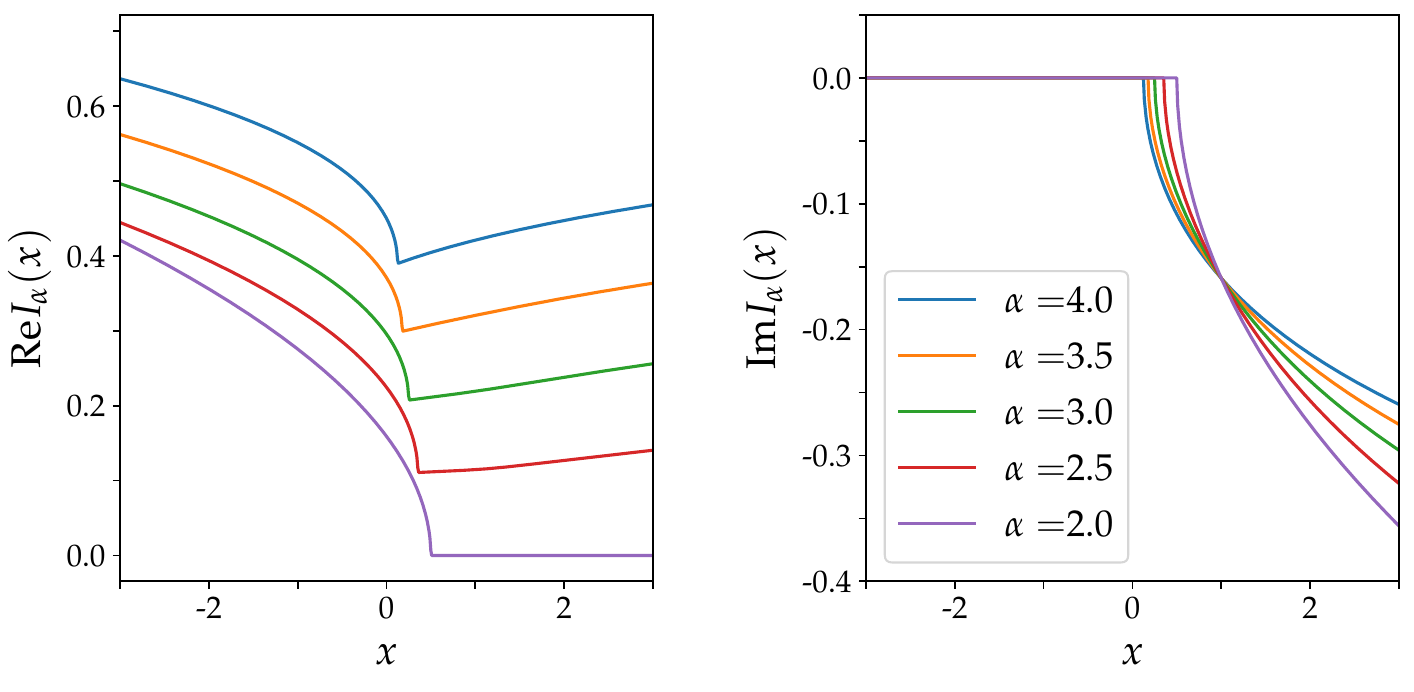}
\caption{Real and imaginary parts of the scaling function $I_\alf(x)$ for
various choices of $\alf$.}
\label{fig:Ialf}
\end{figure}

For $q_t \neq 0$, the particle-hole bubble can be expressed in a scaling form as (see Appendix~\ref{app:A})
\begin{equation} \label{dPi0}
 \delta\Pi_0(q_r,q_t,\om) = \frac{|q_t|}{4v} \, \Big[
 I_\alf\Big( \frac{\om - v q_r}{b|q_t|^\alf} \Big) +
 I_\alf^*\Big( \frac{-\om - v q_r}{b|q_t|^\alf} \Big) \Big] \, ,
\end{equation}
with the dimensionless scaling function
\begin{equation} \label{Ialf}
 I_\alf(x) = \frac{1}{\pi} \int_{-\infty}^{\infty} \frac{d\tk_t}{2\pi}
 \ln\frac{|\tk_t + \frac{1}{2}|^\alf + |\tk_t - \frac{1}{2}|^\alf - x - i0^+}
 {2|\tk_t|^\alf} \, .
\end{equation}
In Fig.~\ref{fig:Ialf} we show $I_\alf(x)$ for various choices of $\alf$.
The integrand in Eq.~\eqref{Ialf} is real for
$x < |\tk_t + \frac{1}{2}|^\alf + |\tk_t - \frac{1}{2}|^\alf$.
For $x > |\tk_t + \frac{1}{2}|^\alf + |\tk_t - \frac{1}{2}|^\alf \geq 2^{1-\alf}$,
the logarithm has a constant imaginary part $-\pi$.
Hence, the imaginary part of $I_\alf(x)$ vanishes for $x < 2^{1-\alf}$, while
\begin{equation}
 \Im I_\alf(x) = - \frac{\tk_t^0(x)}{\pi} \quad \mbox{for} \quad x > 2^{1-\alf} \, ,
\end{equation}
where $\tk_t^0(x)$ is the unique positive solution of the equation
$|\tk_t + \frac{1}{2}|^\alf + |\tk_t - \frac{1}{2}|^\alf = x$.
For $x=1$, one has $\tk_t^0 = \frac{1}{2}$ for any $\alpha$. This is why all the curves in the right panel of Fig.~\ref{fig:Ialf} go through the same crossing point at $x=1$.

For large $|x|$, the scaling function behaves asymptotically as
\begin{equation}
 I_\alf(x) \sim \left\{ \begin{array}{ll}
 \frac{\cot\frac{\pi}{\alf} - i}{2^{\frac{1}{\alf}} \pi} |x|^{\frac{1}{\alf}}
 & \mbox{for} \;\; x \to \infty \, , \\[2mm]
 \frac{(\sin\frac{\pi}{\alf})^{-1}}{2^{\frac{1}{\alf}} \pi} |x|^{\frac{1}{\alf}}
 & \mbox{for} \;\; x \to -\infty \, .
\end{array} \right.
\end{equation}
The next-to-leading correction for large $|x|$ is of order $|x|^{-\frac{1}{\alf}}$.

For $\alf = 2$ and $\alf = 4$, the scaling functions $I_\alf(x)$ can be expressed in terms of square roots,
\begin{eqnarray}
 I_2(x) &=& \frac{1}{\sqrt{2}\pi} \sqrt{\frac{1}{2} - (x + i0^+)} \, ,
 \label{I2x} \\[2mm]
 I_4(x) &=& \frac{1}{\pi} \sqrt{\frac{3}{2} + \sqrt{\frac{1}{4} - 2(x+i0^+)}} \, .
 \label{I4x}
\end{eqnarray}
Inserting $I_2(x)$ into Eq.~\eqref{dPi0} one recovers the well-known expression for the particle-hole bubble for a quadratic dispersion relation. $I_4(x)$ has been computed numerically in Ref.~\cite{debbeler23}, but analytic results were found there only for $x=0$ and $x=\frac{1}{8}$. We derive Eq.~\eqref{I4x} in Appendix~\ref{app:A}.

The RPA effective interaction is given by
\begin{equation} \label{D_rpa}
 D(\bq,iq_0) = \frac{g}{1 + g \chi_0(\bq,iq_0)}
\end{equation}
on the imaginary frequency axis, and by the same expression with $iq_0 \to \om$ on the real frequency axis. At the QCP, $g \chi_0(\bQ,0)$ is equal to minus one, so that
\begin{equation} \label{D_qcp}
 D(\bq,\om) = - \frac{1}{N \delta\Pi_0(\bq,\om)} \, .
\end{equation}
%


\section{Fermion self-energy} \label{sec:self-energy}

To leading order in the effective interaction $D$, the fermion self-energy is given by the one-loop integral
\begin{equation} \label{Sg_rpa}
 \Sg(\bk,ik_0) = - M \int_\bq \int_{q_0} D(\bq,iq_0) \,
 G_0(\bk-\bq,ik_0-iq_0) \, ,
\end{equation}
with $M=1$ for a charge-density and $M=3$ for a spin-density wave instability \cite{sykora18}. This approximation for the self-energy is also known as random-phase approximation (RPA).
Analytic continuation of Eq.~\eqref{Sg_rpa} to the real frequency axis yields \cite{rohe01}
\begin{eqnarray} \label{Sigma_realfreq}
 && \Sg(\bk,\om+i0^+) = - \frac{M}{\pi} \int d\nu \int_\bq \nonumber \\
 && \Big[ b(\nu) \, \Im D(\bq,\nu+i0^+) \, G_0(\bk-\bq,\nu+\om+i0^+) \\[1mm]
 && - \, f(\nu) \, D(\bq,\nu-\om-i0^+) \, \Im G_0(\bk-\bq,\nu+i0^+)
 \Big] \, , \nonumber
\end{eqnarray}
where $b(\nu) = [e^{\beta\nu} - 1]^{-1}$ and $f(\nu) = [e^{\beta\nu} + 1]^{-1}$ are the Bose and Fermi functions, respectively. At zero temperature ($\beta = \infty$) these functions become step functions $b(\nu) = - \Theta(-\nu)$ and $f(\nu) = \Theta(-\nu)$.
In the following we denote $\Sg(\bk,\om+i0^+)$, $G(\bk,\om+i0^+)$, and $D(\bq,\nu+i0^+)$ by $\Sg(\bk,\om)$, $G(\bk,\om)$, and $D(\bq,\nu)$, respectively.

We analyze $\Sg(\bk,\om)$ at the QCP for low frequencies $\om$ and momenta $\bk$ near one of the hot spots on the Fermi surface, which we denote as $\bk_H$. The effective interaction $D(\bq,\om)$ at the QCP is given by Eq.~\eqref{D_qcp} with $\delta\Pi_0(\bq,\om)$ from Eq.~\eqref{dPi0}.
The dominant contributions come from momentum transfers $\bq$ near $\bQ$, so that $\bk-\bq$ is situated near the antipodal hot spot $-\bk_H$. Using relative momentum variables as above, the dispersion relation in the fermion propagator can be expanded as $\xi_{\bk-\bq} = - v (k_r-q_r) + b |k_t-q_t|^\alf$.

To evaluate the self-energy, it is convenient to first consider its imaginary part, and then compute the real part from a Kramers-Kronig relation. The imaginary part of Eq.~\eqref{Sigma_realfreq} reads
\begin{eqnarray} \label{ImSigma1}
 \Im\Sg(\bk,\om) &=& - \frac{M}{\pi} \int d\nu \int_\bq
 \left[ b(\nu) + f(\nu+\om) \right] \, \Im D(\bq,\nu) \nonumber \\
 && \times \, \Im G_0(\bk-\bq,\om+\nu) \, .
\end{eqnarray}
Note that $\Im D(\bq,\nu-i0^+) = - \Im D(\bq,\nu+i0^+)$.
Using the Dirac identity $\Im G_0(\bk,\om) = -\pi \delta(\om-\xi_{\bk})$, the frequency integral in Eq.~\eqref{ImSigma1} can be easily carried out, yielding
\begin{eqnarray} \label{ImSigma2}
 \Im\Sg(\bk,\om) &=& M \int_\bq
 \left[ b(\xi_{\bk-\bq} - \om) + f(\xi_{\bk-\bq}) \right]  \nonumber \\
 &\times& \, \Im D(\bq,\xi_{\bk-\bq} - \om) \, .
\end{eqnarray}
At zero temperature, the sum of Bose and Fermi functions in Eq.~\eqref{ImSigma2} is given by
\begin{equation}
 b(\xi_{\bk-\bq} - \om) + f(\xi_{\bk-\bq}) = \left\{ \begin{array}{rl}
 -1 & \mbox{for} \; 0 < \xi_{\bk-\bq} < \om \, , \\
  1 & \mbox{for} \; \om < \xi_{\bk-\bq} < 0 \, , \\
  0 & \mbox{else} \, , \end{array} \right.
\end{equation}
restricting thus the contributing momentum region.
The integral in Eq.~\eqref{ImSigma2} is convergent even if the momentum integration over $q_r$ and $q_t$ is extended to infinity.

The real part of the self-energy can be obtained from the Kramers-Kronig-type relation
\begin{equation} \label{KK}
 \Sg(\bk,\om) = - \frac{1}{\pi} \int_{-\infty}^{\infty} d\om' \,
 \frac{\Im\Sg(\bk,\om')}{\om - \om' + i0^+} + \mbox{const} \, .
\end{equation}
The last term in this relation is a real constant, which can be absorbed by a shift of the chemical potential. The real part of the self-energy is dominated by contributions from large frequencies in Eq.~\eqref{KK}, where the low-frequency expansion of $\Im\Sg$ is not valid. Since we are not interested in a constant offset but rather in the frequency and momentum dependence of the self-energy near the hot spots, we will analyze the difference
$\delta\Sg(\bk,\om) = \Sg(\bk,\om) - \Sg(\bk_H,0)$, where the leading ultraviolet contributions cancel each other.


\subsection{Frequency dependence at hot spot}

The frequency dependence at the hot spot (for $\bk=\bk_H$) can be derived by a simple rescaling of the integration variables in Eq.~\eqref{ImSigma2}. Substituting $q_r = |\om/v| \tq_r$ and $q_t = |\om/b|^{1/\alf} \tq_t$, one obtains
\begin{equation} \label{ImSigma3}
 \Im\Sg(\bk_H,\om) = - \frac{M}{N} \, A_{s(\om)} |\om| \, ,
\end{equation}
where $A_+$ and $A_-$ are two positive dimensionless numbers depending on $\alf$ and on the sign of $\om$. These numbers are determined by the integral
\begin{equation} \label{Apm}
 A_{s} = - \int'_{\tilde\bq} \! \Im \frac{4s}
 {|\tq_t| \left[ I_\alf \left( \frac{|\tq_t|^\alf - s}{|\tq_t|^\alf} \right) +
 I_\alf^* \left( \frac{-2\tq_r - |\tq_t|^\alf + s}{|\tq_t|^\alf} \right) \right]} \, ,
\end{equation}
where $s = s(\om) = \pm 1$, and the prime at the integral sign indicates a restriction of the integration region to $0 < \tq_r + |\tq_t|^\alf < 1$ for $\om > 0$, and to $-1 < \tq_r + |\tq_t|^\alf < 0$ for $\om < 0$.
Note that the frequency dependence of the self-energy at the hot spot depends neither on $v$ nor on $b$.
In Appendix B we show a plot of the coefficients $A_\pm$ as a function of $\alf$.
They are positive and finite for all $\alf > 2$ and diverge for $\alf \to 2$. The divergence for a quadratic dispersion is due to the vanishing slope of $\delta\Pi_0(q_r,0,0)$ for $q_r < 0$, see Eq.~\eqref{dPi0stat}.

The real part of the self-energy can be obtained from the Kramers-Kronig relation Eq.~\eqref{KK}. With $\Im\Sg(\bk_H,\om)$ as in Eq.~\eqref{ImSigma3}, the integral in Eq.~\eqref{KK} is logarithmically divergent at large frequencies $\om'$. This is due to the fact that the linear frequency dependence has been obtained from an expansion that captures only the asymptotic low frequency behavior, which cannot be extended to all frequencies. The imaginary part of the exact self-energy of any physical system has to vanish in the high-frequency limit. To compute the low-frequency behavior of $\Re\Sg$, we mimic the high-frequency decay of $\Im\Sg$ by imposing an ultraviolet (UV) frequency cutoff $\Lam$, so that the frequency integration in Eq.~\eqref{KK} is restricted to $|\om'| < \Lam$. Defining $\delta\Sg(\bk,\om) = \Sg(\bk,\om) - \Sg(\bk_H,0)$, we then obtain
\begin{equation} \label{ReSg_om}
 \Re \, \delta\Sg(\bk_H,\om) =
 - \frac{M}{N} \, \frac{A_+ + A_-}{\pi} \, \om \ln\frac{\Lam}{|\om|} \, ,
\end{equation}
for $|\om| \ll \Lam$. The imaginary frequency self-energy $\delta\Sg(\bk_H,ik_0)$ is given by the same expression with $\om \mapsto ik_0$.

The logarithm in Eq.~\eqref{ReSg_om} implies a logarithmic divergence of the inverse quasiparticle weight \cite{negele87},
$1 - \partial\Sg(\bk_H,\om)/\partial\om \sim \ln(\Lam/|\om|)$.
Hence, Landau quasiparticles do not exist at the hot spots, and Fermi liquid theory breaks down.

\begin{figure}[tb]
\centering
\includegraphics[width=7cm]{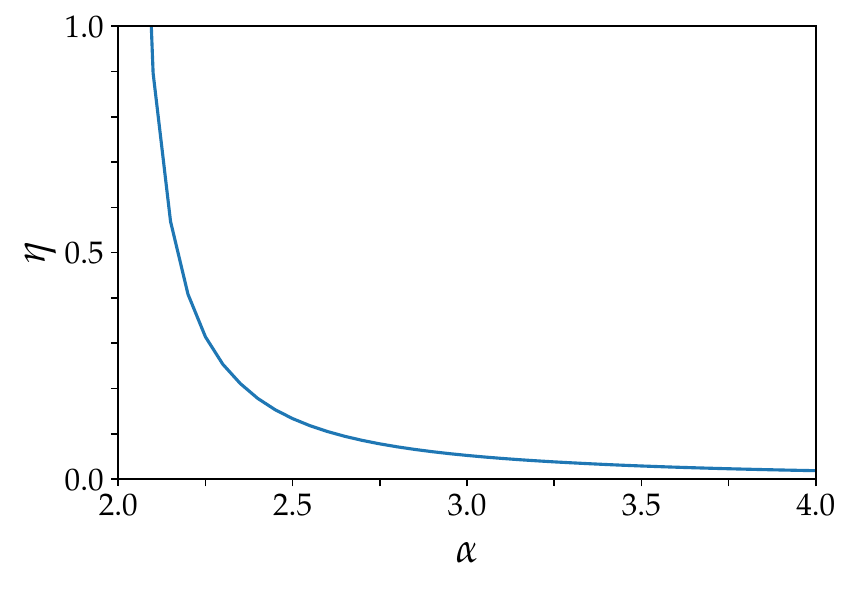}
\caption{Anomalous dimension $\eta$ for $M=1$ and $N=2$ as a function of $\alf$.}
\label{fig:eta}
\end{figure}

Logarithmic divergences are frequently a perturbative manifestation of power-law behavior, especially in (quantum) critical systems. Assuming that the one-loop result in Eq.~\eqref{ReSg_om} reflects the leading order of an expansion of a power-law, one obtains
\begin{equation} \label{Sg_power}
 \om - \delta\Sg(\bk_H,\om) \propto \big( |\om|/\Lam \big)^{-\eta} \om
\end{equation}
at low frequencies, with the anomalous dimension
\begin{equation} \label{eta_om}
 \eta = \frac{M}{N} \, \frac{A_+ + A_-}{\pi} \, .
\end{equation}
Hence, the quasiparticle weight vanishes as $|\om|^{\eta}$ in the low-energy limit.
A plot of $\eta$ as a function of $\alf$ is shown in Fig.~\ref{fig:eta}.
The power-law in Eq.~\eqref{Sg_power} is only an educated guess.
In Sec.~\ref{sec:RG} we will see that $\eta$ is actually scale dependent, so that the quasiparticle weight ultimately vanishes only logarithmically.


\subsection{Frequency and momentum dependencies near hot spot}

We now analyze the momentum and frequency dependence of the self-energy in the vicinity of a hot spot. We consider radial and tangential momentum dependencies separately.

\begin{figure}[tb]
\centering
\includegraphics[width=8.5cm]{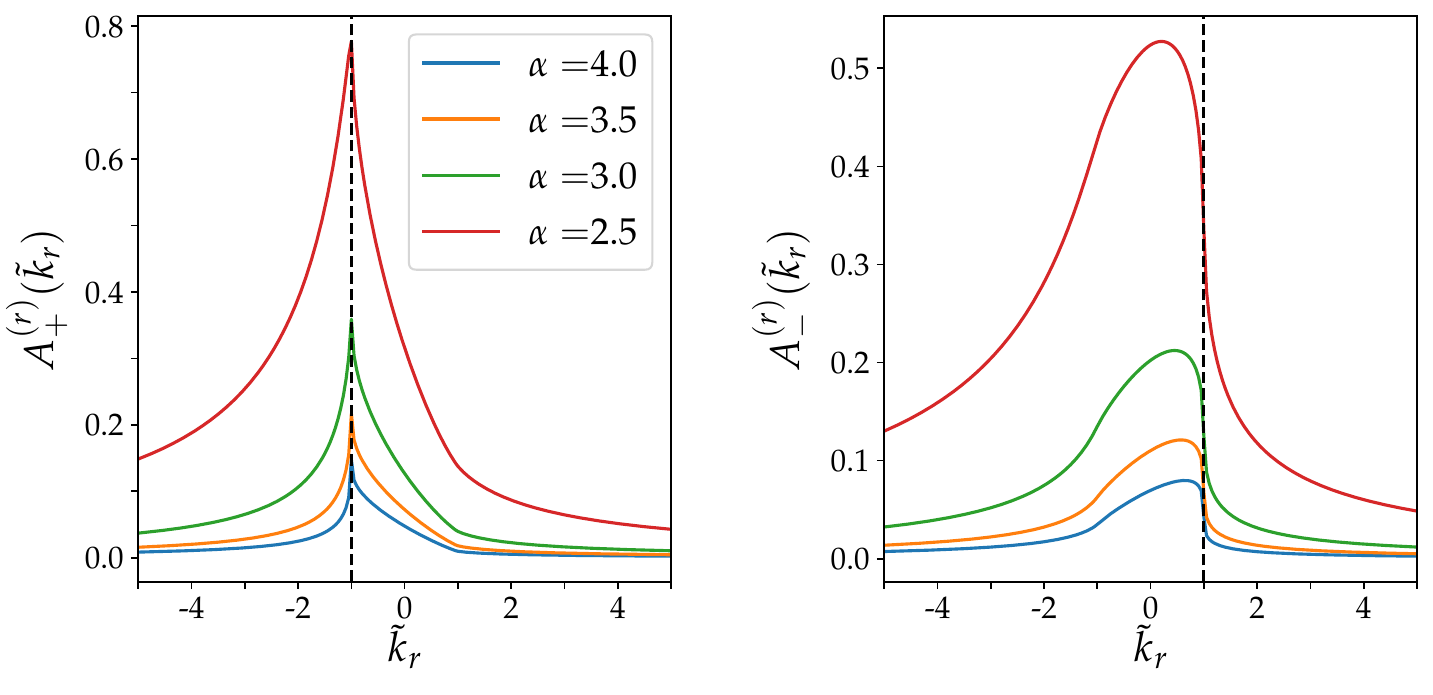}
\caption{Scaling functions $A^{(r)}_\pm(\tk_r)$ for various choices of $\alf$. The vertical dashed lines mark the location of singularities at $\tk_r = \pm 1$.}
\label{fig:Ar}
\end{figure}

For $k_t = 0$, we can express $\Im\Sg(\bk,\om)$ from Eq.~\eqref{ImSigma2} in the scaling form
\begin{equation} \label{ImSigma4}
 \Im\Sg(\bk,\om) = - \frac{M}{N} \, A^{(r)}_{s(\om)}(\tk_r) \, |\om| \, ,
\end{equation}
with the dimensionless scaling functions
%
\begin{align} \label{Apmr}
 & A^{(r)}_{s}(\tk_r) =  	\\
 & - \int'_{\tilde\bq} \Im \frac{4s}
 {|\tq_t| \left[ I_\alf \left( \frac{-\tk_r + |\tq_t|^\alf - s}{|\tq_t|^\alf} \right) +
 I_\alf^* \left( \frac{\tk_r - 2\tq_r - |\tq_t|^\alf + s}{|\tq_t|^\alf} \right) \right]} \, ,
\end{align}
%
where the integration region is restricted to $0 < - \tk_r + \tq_r + |\tq_t|^\alf < 1$ for $\om > 0$, and to $-1 < - \tk_r + \tq_r + |\tq_t|^\alf < 0$ for $\om < 0$.
The rescaled variables are defined by $q_r = |\om/v| \tq_r$, $q_t = |\om/b|^{1/\alf} \tq_t$, and $k_r = |\om/v| \tk_r$.
The scaling functions $A_\pm^{(r)}$ are shown graphically for various choices of $\alf$ in Fig.~\ref{fig:Ar}.
For $\tk_r = 0$ we recover Eq.~\eqref{ImSigma3}, since $A^{(r)}_\pm(0) = A_\pm$ from Eq.~\eqref{Apm}. For small finite $\tk_r$, the leading $\tk_r$ dependence of $A^{(r)}_{s}(\tk_r)$ is linear,
\begin{equation} \label{Apmr_asymp}
 A^{(r)}_\pm(\tk_r) = A_\pm + B_\pm \tk_r + \cO(\tk_r^2) \, ,
\end{equation}
The coefficients $B_\pm$ are shown as functions of $\alf$ in Appendix B.
While $B_+$ is always negative, $B_-$ has a sign change for $\alf \approx 2.3$. Both $B_+$ and $B_-$ diverge for $\alf \to 2$.
For large $|\tk_r|$, $A^{(r)}_{s}(\tk_r)$ decays as $|\tk_r|^{-1}$.

\begin{figure}[tb]
\centering
\includegraphics[width=7cm]{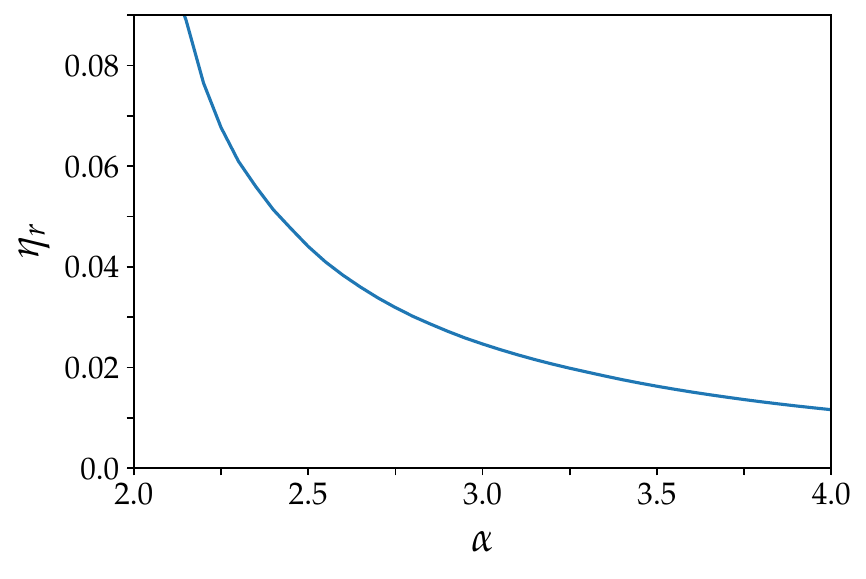}
\caption{Anomalous dimension $\eta_r$ for $M=1$ and $N=2$ as a function of $\alf$.}
\label{fig:eta_r}
\end{figure}

For $k_r \neq 0$ and small $|\om|$, the leading frequency dependence of $\Im\Sg(\bk,\om)$ is quadratic. For $|\om| \gg v |k_r|$, $\Im\Sg(\bk,\om)$ approaches the asymptotic behavior
\begin{equation} \label{ImSigma5}
 \Im\Sg(\bk,\om) \sim - \frac{M}{N} \left[ A_{s(\om)} |\om| + B_{s(\om)} v k_r \right] \, ,
\end{equation}
which follows from Eq.~\eqref{Apmr_asymp}.
Inserting this asymptotic dependence into the Kramers-Kronig relation Eq.~\eqref{KK} with an ultraviolet frequency cutoff $\Lam$, one obtains the leading $k_r$ dependence of the real part of the self-energy at zero frequency as
\begin{equation} \label{ReSigmakr}
 \delta\Sg(\bk,0) \sim \, \frac{M}{N} \, \frac{B_- - B_+}{\pi} \, v k_r \,
 \ln\frac{\Lam}{v |k_r|}  \, .
\end{equation}
The same result is obtained by inserting Eq.~\eqref{ImSigma4} with the full (not expanded) scaling function $A^{(r)}_s(\tk_r)$ into the Kramers-Kronig integral.
The difference $B_- - B_+$ in Eq.~\eqref{ReSigmakr} is always positive (for all $\alf$), and diverges for $\alf \to 2$.
Assuming, as before, that the logarithm reflects the leading contribution from a power-law, we might expect a momentum dependence of the form
\begin{equation} \label{Sg_power2}
 v k_r + \delta\Sg(\bk,0) \propto \big( v |k_r|/\Lam \big)^{-\eta_r} v k_r
\end{equation}
for small $k_r$, with the anomalous dimension
\begin{equation} \label{eta_r}
 \eta_r = \frac{M}{N} \, \frac{B_- - B_+}{\pi} \, .
\end{equation}
We show $\eta_r$ for $M=1$ and $N=2$ as a function of $\alf$ in Fig.~\ref{fig:eta_r}.
The effective Fermi velocity \cite{negele87} given by
$v(k_r) = \left( 1 - \partial\Sg/\partial\om \right)^{-1}
 \left( 1 + \partial\Sg/\partial k_r \right) v$ is proportional to
$|\om|^\eta |k_r|^{-\eta_r}$ with $\om = v k_r$, and thus
$\bar v(k_r) \propto |k_r|^{\eta - \eta_r}$.
This quantity vanishes for $k_r \to 0$, since $\eta > \eta_r$ for all $\alf$.
However, the renormalization group analysis in Sec.~\ref{sec:RG} shows that $v(k_r)$ actually vanishes only logarithmically.

\begin{figure}[tb]
\centering
\includegraphics[width=8.5cm]{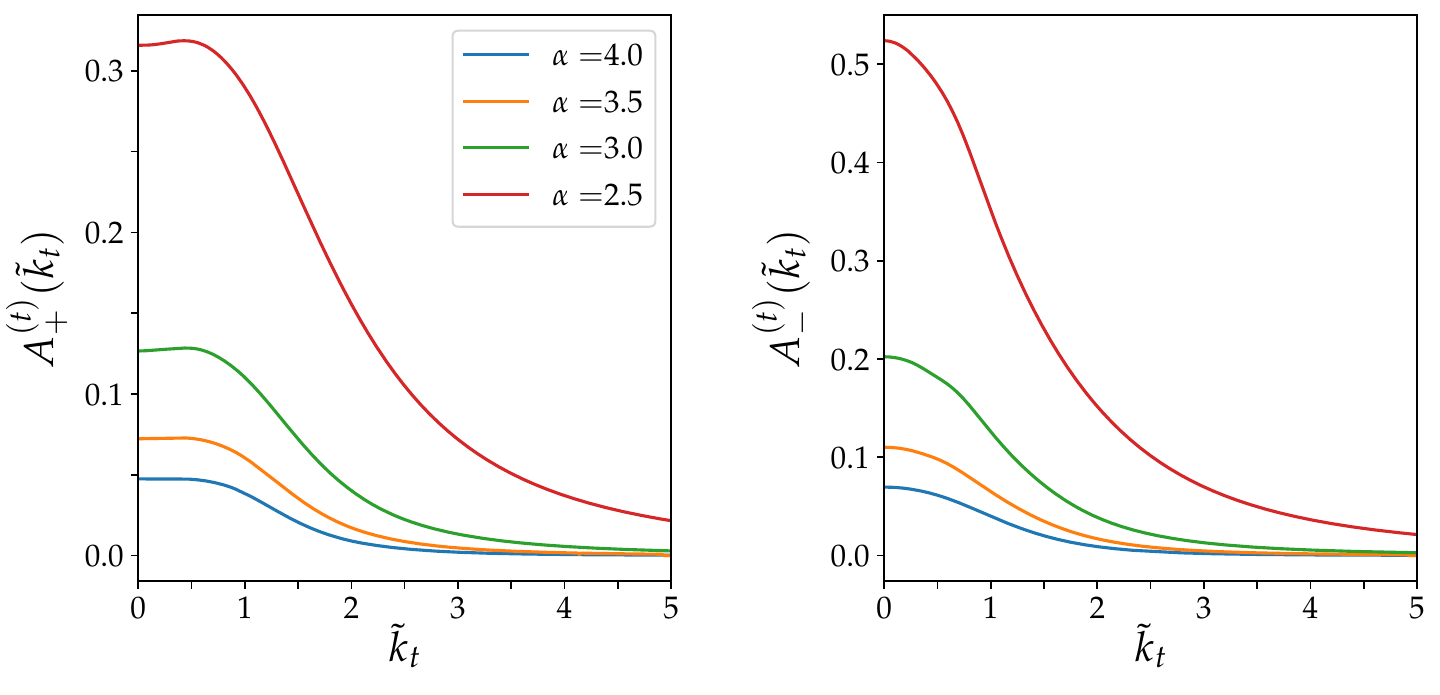}
\caption{Scaling functions $A^{(t)}_\pm(\tk_t)$ for various choices of $\alf$.}
\label{fig:At}
\end{figure}

We now discuss the tangential momentum dependence of the self-energy.
For $k_r = 0$, we can express $\Im\Sg(\bk,\om)$ in the scaling form
\begin{equation} \label{ImSigma6}
 \Im\Sg(\bk,\om) = - \frac{M}{N} \, A^{(t)}_{s(\om)}(\tk_t) \, |\om| \, ,
\end{equation}
with the dimensionless scaling functions
%
\begin{align} \label{Apmt}
 & A^{(t)}_s(\tk_t) = \\
 & - \int'_{\tilde\bq} \!\! \Im \frac{4s}
 {|\tq_t| \left[ I_\alf \left( \frac{|\tk_t-\tq_t|^\alf - s}{|\tq_t|^\alf} \right) +
 I_\alf^* \left( \frac{- 2\tq_r - |\tk_t-\tq_t|^\alf + s}{|\tq_t|^\alf} \right) \right]} \, ,
\end{align}
%
where the integration region is restricted to $0 < \tq_r + |\tk_t-\tq_t|^\alf < 1$ for $\om > 0$, and to $-1 < \tq_r + |\tk_t-\tq_t|^\alf < 0$ for $\om < 0$.
The rescaled variables are defined by $q_r = |\om/v| \tq_r$, $q_t = |\om/b|^{1/\alf} \tq_t$, and $k_t = |\om/b|^{1/\alf} \tk_t$.
The scaling functions $A^{(t)}_\pm(\tk_t)$ are symmetric under $\tk_t \mapsto -\tk_t$.
Their behavior as a function of (positive) $\tk_t$ is shown for various choices of $\alf$ in Fig.~\ref{fig:At} \cite{footnote_At}.
A numerical analysis shows that, to quartic order, $A^{(t)}_\pm(\tk_t)$ can be expanded as
\begin{equation} \label{Atexp}
 A^{(t)}_\pm(\tk_t) = A_\pm + C_\pm \tk_t^2 + D_\pm \tk_t^4 + \cO(\tk_t^6)
\end{equation}
for small $\tk_t$. At least to that order no non-integer powers of $\tk_t$, such as $|\tk_t|^\alf$, contribute. Plots of the coefficients $C_\pm$ and $D_\pm$ as functions of $\alf$ are shown in Appendix B. For large $|\tk_t|$, $A^{(t)}_s(\tk_t)$ decays as $|\tk_t|^{-\alf}$.

Eqs.~\eqref{ImSigma6} and \eqref{Atexp} imply that $\Im\Sg(\bk,\om)$ behaves as
\begin{eqnarray} \label{ImSigma7}
 \Im\Sg(\bk,\om) &\sim& - \frac{M}{N} \Big[
 A_{s(\om)} |\om| + C_{s(\om)} b^\frac{2}{\alf} |\om|^{1-\frac{2}{\alf}} k_t^2
 \nonumber \\
 &+& D_{s(\om)} b^\frac{4}{\alf} |\om|^{1-\frac{4}{\alf}} k_t^4 \Big]
\end{eqnarray}
for $|\om| \gg b |k_t|^\alf$.
Computing $\delta\Sg(\bk,0) = \Sg(\bk,0) - \Sg(\bk_H,0)$ from the Kramers-Kronig relation Eq.~\eqref{KK}, the first term in Eq.~\eqref{ImSigma7} cancels, but the second one generates an ultraviolet divergent term proportional to
$(C_- - C_+) \Lam^{1-2/\alf} k_t^2$ for any $\alf > 2$.
This term, along with generic regular many-body contributions, leads to a renormalized dispersion relation $\bar\xi_\bk$ with a quadratic tangential momentum dependence, in conflict with our original assumption. However, the case of a dispersion with a vanishing quadratic dependence on $k_t$ can be restored by adding a quadratic contribution to the bare dispersion as a counterterm, which cancels the quadratic self-energy correction.

\begin{figure}[tb]
\centering
\includegraphics[width=7cm]{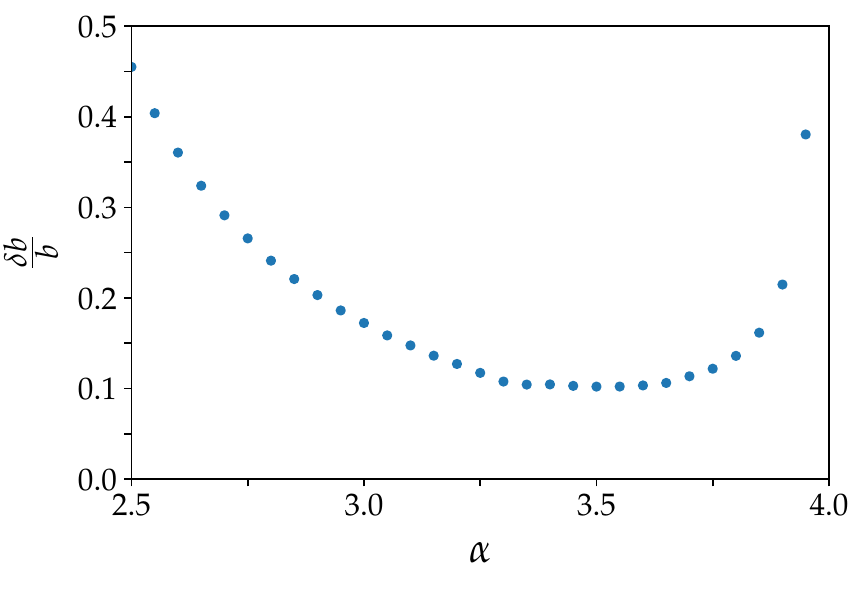}
\caption{$\delta b/b$ for $M=1$ and $N=2$ as a function of $\alf$.}
\label{fig:deltab}
\end{figure}

To compute the $k_t$ dependence of the remaining contributions to $\delta\Sg(\bk,0)$, we subtract also the quadratic part from $\Im\Sg(\bk,\om)$ in Eq.~\eqref{ImSigma7}, that is, we compute the Kramers-Kronig integral for
$\delta \Im\Sg(\bk,\om) = \Im\Sg(\bk,\om) + \frac{M}{N} \big[
 A_{s(\om)} |\om| + C_{s(\om)} b^{2/\alf} |\om|^{1-2/\alf} k_t^2 \big]$.
For $2 < \alf < 4$ this integral is convergent and the result can be written as
\begin{equation} \label{deltaSgt}
 \delta\Sg(\bk,0) = \delta b \, |k_t|^\alf \, ,
\end{equation}
where
\begin{equation} \label{deltab}
 \delta b = b \, \frac{M}{N\pi} \int_0^\infty d\tom \left[
 \delta A_-^{(t)}(\tom^{-\frac{1}{\alf}}) - \delta A_+^{(t)}(\tom^{-\frac{1}{\alf}})
 \right] \, ,
\end{equation}
where $\delta A_s^{(t)}(\tk_t) = A_s^{(t)}(\tk_t) - A_s - C_s \tk_t^2$.
The integral in Eq.~\eqref{deltab} converges in the infrared (small $\tom$) for any $\alf > 2$ and in the ultraviolet (large $\tom$) for $\alf < 4$.
The $\alf$ dependence of $\delta b/b$ for $2 < \alf < 4$ is shown in Fig.~\ref{fig:deltab}.
For $\alf = 4$ the integral diverges logarithmically in the ultraviolet, and one obtains the result presented already in Ref.~\cite{debbeler23},
\begin{equation}
 \delta\Sg(\bk,0) \sim \frac{M}{\pi N} (D_{-} - D_{+}) \,
 bk_t^4 \ln\frac{\Lam}{bk_t^4}
\end{equation}
for $bk_t^4 \ll \Lam$, which suggests a power-law with the anomalous dimension \cite{debbeler23}
\begin{equation} \label{eta_t}
 \eta_t = \frac{M}{N} \frac{D_{-} - D_{+}}{\pi} \, .
\end{equation}
The renormalization group analysis in the next section reveals that $\eta_t$ is scale dependent, so that ultimately only logarithmic corrections to the quartic tangential momentum dependence are obtained.
The logarithmic UV divergence of $\delta b$ for $\alf = 4$ is reflected by a divergence of $\delta b$ for $\alf < 4$ in the limit $\alf \to 4$:
\begin{equation} \label{epspole}
 \delta b \sim \frac{4}{\eps} \, \eta_t \, b \, ,
\end{equation}
for $\eps \to 0$, with $\eps = 4 - \alf$ and $\eta_t$ from Eq.~\eqref{eta_t}.


\section{Renormalization group analysis} \label{sec:RG}

The logarithmic divergences encountered in the one-loop self-energy imply a breakdown of perturbation theory and indicate possible power law behavior with anomalous exponents. For a controlled treatment of these divergences, we now use the field theoretic renormalization group \cite{dicastro76,amit84}.

For a renormalization group analysis it is convenient to describe our system by a quantum field theory with fermion and boson fields, where the latter represent fluctuations of the order parameter. The corresponding action has the form
\begin{eqnarray} \label{action1}
 {\cal S} &=& - \int_k \sum_\sg G_0^{-1}(k) \psi_\sg^*(k) \psi_\sg(k)
 - \frac{1}{2} \int_q \! D^{-1}(q) \phi^*(q) \phi(q) \nonumber \\
 && + \, u \int_{k,q} \sum_\sg \phi(q) \psi_\sg^*(k+q) \psi_\sg(k) \, ,
\end{eqnarray}
where $\psi_\sg^*(k)$ and $\psi_\sg(k)$ with $\sg \in \{ 1,\dots,N \}$ are fermionic fields corresponding to fermionic creation and annihilation operators, respectively, while $\phi(q)$ is a bosonic field describing the order parameter fluctuations. The variables $k = (\bk,ik_0)$ and $q = (\bq,iq_0)$ contain frequencies and momenta; $\int_k$ and $\int_q$ are shorthand notations for $\int_\bk \int_{k_0}$ and $\int_\bq \int_{q_0}$, respectively.
Note that the interaction term in Eq.~\eqref{action1} is hermitian since $\phi(-q) = \phi^*(q)$.

An action of the form Eq.~\eqref{action1} can be derived from the original purely fermionic action by decoupling the two-fermion interaction via a Hubbard-Stratonovich transformation \cite{negele87}. The boson propagator is thereby obtained as $D(q) = D_0(q) = g$, and the bare Yukawa coupling is $u = 1$.
In one-loop approximation the bosonic self-energy is given by $u^2 N \Pi_0(q)$, so that the boson propagator becomes $D(q) = \left[ g^{-1} - N \Pi_0(q) \right]^{-1}$, which is equal to the RPA effective interaction in Eq.~\eqref{D_rpa}.
At the quantum critical point, where $N \Pi_0(\bQ,0) = g^{-1}$, we thus obtain $D^{-1}(q) = - N \delta\Pi_0(q)$.
Our scaling and renormalization group analysis is based on the action Eq.~\eqref{action1} with
$D^{-1}(q) = - N \delta\Pi_0(q)$. The one-loop boson self-energy has thus already been taken into account. An analogous procedure has been used previously for a renormalization group analysis of the Ising nematic quantum critical point \cite{metlitski10_nem,holder15}.

Since singular contributions are due to fermions with momenta near the two hot spots $\pm\bk_H$, we introduce relative momentum coordinates and expand around the hot spots as in the perturbative analysis in the preceding section. We then need to label fermion fields with an additional index $l = \pm$ to distinguish fields with momenta near $\bk_H$ from those near $-\bk_H$, and the bosonic fields are labeled by their momentum transfer being close to $\bQ$ or $-\bQ$ as $\phi(\pm\bQ+\bq,q_0) = \phi_\pm(\bq,q_0) = \phi_\pm(q)$.
The action Eq.~\eqref{action1} can then be written as
\begin{eqnarray} \label{action2}
 {\cal S} &=& \int_k \sum_{l,\sg} \left(-ik_0 + vk_r + b|k_t|^\alf \right)
 \psi_{l,\sg}^*(k) \psi_{l,\sg}(k)^{\phantom *} \nonumber \\
 &+& N \int_q \delta\Pi_0(q) \phi_{+}(q) \phi_{-}(-q) \nonumber \\
 &+& u \int_{k,q} \sum_{l,\sg}
 \phi_l(q) \psi_{l,\sg}^*(k+q) \psi_{-l,\sg}^{\phantom *}(k) \, .
\end{eqnarray}

We start with a simple analysis of canonical scaling dimensions. Rescaling fermionic momenta and frequencies as
\begin{equation}
 k_0 \to s k_0 \, , \; k_r \to s k_r \, , \; k_t \to s^{1/\alf} k_t \, ,
\end{equation}
and anagously for bosonic momenta and frequencies $q_0$, $q_r$, and $q_t$, implies
\begin{equation}
 G_0(k) \to s^{-1} G_0(k) \, , \; D(q) \to s^{-1/\alf} D(q) \, .
\end{equation}
The latter relation follows from Eq.~\eqref{dPi0}. Scaling frequencies by a linear factor $s$, the scaling of the momentum variables is determined by requiring a homogeneous scaling relation for $G_0(k)$, that is, all terms contributing to $G_0^{-1}(k)$ should scale linearly with $s$.
Requiring that the quadratic parts of the action be scale invariant, and taking into account that the integration measures in $\int_k$ and $\int_q$ scale as $s^{2+1/\alf}$, the fields need to be rescaled as
%
\begin{align}
 & \psi_{l,\sg} \to s^{-\frac{3}{2}-\frac{1}{2\alf}} \psi_{l,\sg} \, , \;
 \psi_{l,\sg}^* \to s^{-\frac{3}{2}-\frac{1}{2\alf}} \psi_{l,\sg}^* \, , \\
 & \phi_l \to s^{-1-\frac{1}{\alf}} \phi_l \, .
\end{align}
%
Inserting this scaling behavior of the fields into the interaction term in Eq.~\eqref{action2}, one finds that this term is scale invariant, so that the Yukawa coupling constant $u$ is marginal.

We now turn to the renormalization group analysis with fluctuations on one-loop level, which corresponds to the leading order in a $1/N$ expansion. While the bosonic one-loop self-energy is finite, the fermionic self-energy exhibits logarithmic divergences which were derived in the preceding section. Expressing the self-energy as a function of the three variables $k_r$, $k_t$, and $k_0$, and subtracting $\Sg(\bk_H,0)$, we found
\begin{eqnarray}
 \delta\Sg(0,0,ik_0) &=& - \eta u^2 \, ik_0 \ln\frac{\Lam}{|k_0|} \, , \\
 \delta\Sg(k_r,0,0) &=& \eta_r u^2 \, v k_r \ln\frac{\Lam}{v|k_r|} \, ,
\end{eqnarray}
for any $\alf > 2$, where $\eta$ and $\eta_r$ are the (tentative) anomalous dimensions defined in Eqs.~\eqref{eta_om} and \eqref{eta_r}, respectively. Note that we have explicitly included the coupling constant $u = 1$, for reasons that will become clear below.
For $\alf = 4$ there is also a logarithmic divergence in the tangential momentum dependence,
\begin{equation}
\delta\Sg(0,k_t,0) = \eta_t u^2 \, b k_t^4 \ln\frac{\Lam}{bk_t^4} \, ,
\end{equation}
with $\eta_t$ from Eq.~\eqref{eta_t}, while for $\alf < 4$ only a finite correction of the order $|k_t|^{\alf}$ is obtained. However, the prefactor $\delta b$ of that correction diverges for $\alf \to 4$, see Eq.~\eqref{epspole}. For $\alf > 4$, the coefficient of the quartic contribution to the self-energy diverges as a power of the ultraviolet cutoff. In the following we focus on the cases $2 < \alf \leq 4$.

A complete one-loop calculation for an action of the form Eq.~\eqref{action1} involves also a one-loop correction of the Yukawa vertex. However, for ordering wave vectors distinct from half a reciprocal lattice vector, there is no choice of momenta in the vertex correction at which the singularities of the propagators coalesce \cite{altshuler95}, so that the vertex correction is finite. Hence, the only divergences come from the fermion self-energy.

Since the leading tangential momentum dependence of the self-energy diverges for $\alf = 4$, while it is finite for $\alf < 4$, we need to distinguish these two cases.


\subsection{$\boldsymbol{\alf = 4}$}

Following the field theoretic renormalization procedure \cite{dicastro76,amit84}, we define a renormalized fermionic two-point vertex function
\begin{equation}
 \bar\Gam^{(2)}(k;\bar v, \bar b, \bar u; \lam) =
 Z(u;\lam/\Lam) \, \Gam^{(2)}(k;v,b,u;\Lam) \, ,
\end{equation}
where $\Gam^{(2)} = G_0^{-1} - \delta\Sg$ is the unrenormalized vertex function,
\begin{equation} \label{Z}
 Z = 1 - \eta u^2 \ln\frac{\Lam}{\lam} \, ,
\end{equation}
and the renormalized parameters $\bar u$, $\bar v$, $\bar b$ are defined as
\begin{eqnarray}
 \bar u &=& Z u \, , \label{bar_u} \\
 \bar v &=& Z \left( 1 + \eta_r u^2 \ln\frac{\Lam}{\lam} \right) v \, , \label{bar_v} \\
 \bar b &=& Z \left( 1 + \eta_t u^2 \ln\frac{\Lam}{\lam} \right) b \, , \quad
 \mbox{for} \quad \alf = 4 \, . \label{bar_b}
\end{eqnarray}
The energy scale $\lam$ can be chosen arbitrarily, or it may be used to satisfy normalization conditions of the renormalized theory at specific frequencies and momenta \cite{dicastro76, amit84}.
To second order in the coupling constant, the renormalized two-point vertex, expressed as a function of the renormalized parameters $\bar v$, $\bar b$, and $\bar u$, is finite for $\Lam \to \infty$:
\begin{eqnarray}
 && \bar\Gam^{(2)}(k;\bar v, \bar b, \bar u; \lam) =
 ik_0 \left( 1 + \eta \bar u^2 \ln\frac{\lam}{|k_0|} \right) \nonumber \\
 && - \bar v k_r \left( 1 + \eta_r \bar u^2 \ln\frac{\lam}{\bar v|k_r|} \right)
 - \bar b k_t^4 \left( 1 + \eta_t \bar u^2 \ln\frac{\lam}{\bar bk_t^4} \right) .
 \hskip 7mm
\end{eqnarray}

The renormalization of the Yukawa coupling, Eq.~\eqref{bar_u}, is determined by the renormalization of the Yukawa vertex
\begin{equation}
 \bar\Gam^{(2,1)}(k,q;\bar u;\lam) = Z(u;\lam/\Lam) \, \Gam^{(2,1)}(k,q;u;\Lam) \, .
\end{equation}
Since $\Gam^{(2,1)}$ is finite at the one-loop level, the renormalization of $u$ is determined exclusively by $Z$. Discarding non-universal finite one-loop vertex corrections, we approximate $\Gam^{(2,1)}(0,0;u;\Lam)$ by its bare value $u$, so that $\bar u = \bar\Gam^{(2,1)}(0,0;\bar u;\lam) = Zu$.

The renormalization group flow is obtained from the evolution of $\bar\Gam^{(2)}(k;\bar v, \bar b, \bar u; \lam)$ upon varying the energy scale $\lam$ \cite{dicastro76, amit84}. The bare vertex $\Gam^{(2)}(k;v,b,u;\Lam)$ does not depend on $\lam$, so that
\begin{equation}
 \left. \lam \frac{d}{d\lam} Z^{-1}
 \bar\Gam^{(2)}(k;\bar v, \bar b, \bar u; \lam) \right|_{v,b,u,\Lam} = 0 \, ,
\end{equation}
and thus
\begin{equation} \label{floweq_Gamma2}
 \big[ \lam \partial_\lam + \beta_{\bar u} \partial_{\bar u} +
 \beta_{\bar v} \partial_{\bar v} + \beta_{\bar b} \partial_{\bar b}
    - \gam \big] \bar\Gam^{(2)}(k;\bar v,\bar b,\bar u;\lam) = 0 \, , \;
\end{equation}
where $\partial_\lam$ is a partial derivative with respect to $\lam$,
\begin{equation}
 \beta_{\bar x} =
 \left. \lam \frac{\partial\bar x}{\partial\lam} \right|_{v,b,u,\Lam}
\end{equation}
for $\bar x = \bar v$, $\bar b$, $\bar u$, and
\begin{equation} \label{gamma}
 \gam =  \left. \lam \frac{\partial \ln Z}{\partial\lam} \right|_{v,b,u,\Lam} \, .
\end{equation}
Inserting Eq.~\eqref{Z} into Eq.~\eqref{bar_u}, we obtain
\begin{equation} \label{beta_u}
 \beta_{\bar u} = \eta {\bar u}^3 \, .
\end{equation}
On the right hand side we have replaced $u$ by $\bar u$, which is justified since $\bar u - u$ is of order $(\bar u)^3$. The flow equation for $\bar u$ does not depend on the other variables and can be easily integrated. With the initial condition $\bar u = \bar u_0 = 1$ for $\lam = \lam_0$ we find
\begin{equation} \label{bar_u_sol}
 \bar u = \frac{1}{\sqrt{1 + 2\eta \ln(\lam_0/\lam)}} \, .
\end{equation}
Hence, $\bar u$ tends to $\bar u^* = 0$ for $\lam \to 0$, albeit very slowly.
Inserting Eq.~\eqref{Z} into Eq.~\eqref{gamma} yields, to order $\bar u^2$,
\begin{equation}
 \gamma = \eta {\bar u}^2 =
 \frac{\eta}{1 + 2\eta \ln(\lam_0/\lam)} \, .
\end{equation}
Eqs.~\eqref{bar_v} and \eqref{bar_b} yield, respectively,
\begin{eqnarray}
 \beta_{\bar v} &=& (\eta - \eta_r) \, {\bar u}^2 {\bar v} \, , \label{beta_v} \\
 \beta_{\bar b} &=& (\eta - \eta_t) \, {\bar u}^2 {\bar b} \, . \label{beta_b}
\end{eqnarray}
Inserting the solution for $\bar u$, Eq.~\eqref{bar_u_sol}, the flow equations for $\bar v$ and $\bar b$ can be easily integrated to
\begin{eqnarray}
 \bar v &=& \bar v_0 \big[ 1 +
 2\eta \ln(\lam_0/\lam) \big]^{\frac{\eta_r - \eta}{2\eta}} \, ,
 \label{bar_v_sol} \\
 \bar b &=& \bar b_0 \big[ 1 +
 2\eta \ln(\lam_0/\lam) \big]^{\frac{\eta_t - \eta}{2\eta}} \, ,
 \label{bar_b_sol}
\end{eqnarray}
where $\bar v_0$ and $\bar b_0$ are the initial values of $\bar v$ and $\bar b$, repectively, at $\lam = \lam_0$.
Since $\eta_r$ and $\eta_t$ are both smaller than $\eta$, the renormalized quantities $\bar v$ and $\bar b$ decrease upon decreasing $\lam$. Initially this decrease is very slow, while ultimately (for $\lam \ll \lam_0 e^{-1/(2\eta)}$) they vanish as some power of a logarithm.

From the flow of the renormalized quantities and $\gamma$ as a function of $\lam$ we can obtain the momentum and frequency dependences of the fermionic two-point vertex $\Gamma^{(2)}$, following the standard procedure \cite{dicastro76,amit84}.
We begin with the frequency dependence at the hot spot. For $k_r = k_t = 0$, the flow equation \eqref{floweq_Gamma2} for the renormalized vertex $\bar\Gamma^{(2)}$ can be integrated to
\begin{equation}
 \bar\Gamma^{(2)}(0,0,ik_0) =
 ik_0 \big[ 1 + 2\eta \ln(\lam/|k_0|) \big]^\frac{1}{2} \, ,
\end{equation}
with the normalization condition $\bar\Gamma^{(2)}(0,0,ik_0) = ik_0$ for $|k_0| = \lam$. A simple dimensional argument then yields
\begin{equation}
 \Gamma^{(2)}(0,0,ik_0) = ik_0 \big[ 1 + 2\eta \ln(\Lam/|k_0|) \big]^\frac{1}{2} \, .
\end{equation}
This result agrees with the perturbative expression to linear order in $\eta$. However, the naive expectation of a power law with an anomalous exponent $\eta$ was not confirmed by the renormalization group analysis. Instead, only a weaker logarithmic correction to the bare frequency dependence is obtained.
This behavior is similar to the momentum dependence of the two-point vertex in the $\phi^4$-theory at the critical dimension $d_c = 4$, where the interaction is also marginally irrelevant and leads to logarithmic corrections of the momentum dependence \cite{amit84}. In that case, however, the beta function is quadratic in the coupling to leading order, while in our model it is cubic, so that the renormalized coupling vanishes more slowly.

From the flow of the renormalized quantities $\bar v$ and $\bar b$ in Eqs.~\eqref{bar_v_sol} and \eqref{bar_b_sol}, respectively, we obtain momentum dependent coefficients of the fluctuation-corrected dispersion as
\begin{eqnarray}
 v(k_r) &=& v \left( 1 +
 2\eta \ln\frac{\Lam}{v|k_r|} \right)^{\frac{\eta_r - \eta}{2\eta}} \, ,
 \label{v_sol} \\[2mm]
 b(k_t) &=& b \left( 1 +
 2\eta \ln\frac{\Lam}{bk_t^4} \right)^{\frac{\eta_t - \eta}{2\eta}} \, ,
 \label{b_sol}
\end{eqnarray}
via dimensional arguments. These results agree with the perturbative expressions to leading order in the anomalous dimensions $\eta$, $\eta_r$, and $\eta_t$. Since $\eta$ is larger than $\eta_r$ and $\eta_t$, both $v(k_r)$ and $b(k_t)$ vanish upon approaching the hot spot in momentum space, albeit very slowly.


\subsection{$\boldsymbol{\alf < 4}$}

For $\alf < 4$ the tangential momentum dependence of the fermion propagator acquires only a finite correction, as given by $\delta b$ in Eq.~\eqref{deltab}. The divergences in the self-energy corrections to the frequency and radial momentum dependence are qualitatively the same as for $\alf = 4$. These divergences can therefore be treated by the same renormalization group procedure as for $\alf = 4$. The defining equations of $Z$, $\bar u$, and $\bar v$, as well as their flow equations, remain the same, and one obtains the same results for the frequency and radial momentum dependence as for $\alf = 4$. The dependence on $\alf$ enters only via the parameters $\eta$ and $\eta_r$. As to the tangential momentum dependence, a renormalized coefficient $\bar b$ could be defined as
\begin{equation} \label{bar_b2}
 \bar b = Z \left( b + u^2 \delta b \right) \, .
\end{equation}
The flow of $\bar b$ is then driven only by the divergence of the $Z$-factor.

The above procedure is satisfactory as long as $\delta b/b$ is small, which is the case for $\alf$ staying sufficiently far away from two and four. For $\alf \to 2$, not only $\delta b/b$ becomes large, but also $\eta$ and $\eta_r$, so that the one-loop theory breaks down completely. For $\alf \to 4$, however, the frequency and radial momentum dependence seem to be well captured by the one-loop approximation, and we can deal with the tangential momentum dependence by resorting to an $\eps$-expansion in $\eps = 4 - \alf$.

To this end, we define our renormalization group analysis for a scaling of the tangential momentum variable as $k_t \to s^{1/4} k_t$ instead of $k_t \to s^{1/\alpha} k_t$. In this way, the coefficient $b$ becomes a relevant variable for $\alf < 4$ already at tree level (without loop corrections), while the Yukawa coupling $u$ remains marginal.
The self-energy from Eq.~\eqref{deltaSgt} is written accordingly in the form
\begin{equation}
 \delta\Sg(0,k_t,0) = \frac{4}{\eps} \eta_t u^2 b |k_t|^{-\eps} k_t^4 \, ,
\end{equation}
where $\eta_t$ can be evaluated for $\alf=4$ to leading order in $\eps$.
The renormalized vertex $\bar\Gam^{(2)} = Z \Gam^{(2)}$ is finite for $\Lam \to \infty$ and $\eps \to 0$ if $\bar u$ and $\bar v$ are chosen as before, and
\begin{equation}
 \bar b = Z \left( 1 + \frac{4}{\eps} \eta_t u^2 \right) \lam^{-\eps/4} \, b \, .
\end{equation}
Applying $\lam\partial_\lam$ to $\bar b$, and expanding in $\bar u$ and $\eps$, we obtain the $\beta$-function for the flow of $\bar b$ to order $\bar u^2$ and order $\eps$,
\begin{equation}
 \beta_{\bar b} = - \frac{\eps}{4} \, \bar b +
 (\eta - \eta_t) \bar u^2 \, \bar b \, .
\end{equation}
For $\eps \to 0$ this $\beta$-function reduces continuously to the $\beta$-function for $\alf=4$ in Eq.~\eqref{beta_b}.
The flow equation for $\bar b$ can be integrated to
\begin{equation} \label{bar_b_sol2}
 \bar b = \bar b_0 (\lam_0/\lam)^{\eps/4}
 \big[ 1 + 2\eta\ln(\lam_0/\lam) \big]^\frac{\eta_t - \eta}{2\eta} \, ,
\end{equation}
where $\bar b_0$ is the initial value of $\bar b$ at $\lam = \lam_0$.
Using once again the common dimensional arguments \cite{amit84}, we thus obtain the tangential momentum dependence of the fluctuation corrected dispersion in the form $b(k_t) |k_t|^\alf$, with
\begin{equation}
 b(k_t) = b \left( 1 +
 2\eta \ln\frac{\Lam}{b|k_t|^\alf} \right)^{\frac{\eta_t - \eta}{2\eta}} \, .
\end{equation}
The first scale dependent factor in Eq.~\eqref{bar_b_sol2} simply shifts the exponent from $k_t^4$ to $|k_t|^\alf$, while the second factor leads to a logarithmic correction. For $\alf \to 4$ we recover the behavior of $b(k_t)$ in Eq.~\eqref{b_sol}.

Since $\eta_r > \eta_t$ for all $\alf$, the renormalized Fermi velocity $v(k_r)$ vanishes slightly faster than $b(k_t)$. Hence, the renormalized Fermi surface is slightly (logarithmically) flatter than the bare one.


\section{Conclusion} \label{sec:Conclusion}

We have analyzed quantum fluctuation effects at the onset of incommensurate $2k_F$ charge- or spin-density wave order in two-dimensional metals, for a model where the ordering wave vector $\bQ$ connects a single pair of hot spots on the Fermi surface with vanishing Fermi surface curvature. The tangential momentum dependence of the bare dispersion near the hot spots is proportional to $|k_t|^\alf$ with $\alf > 2$.
Varying $\alf$ yields a smooth interpolation between the conventional parabolic case with a finite Fermi surface curvature, for $\alf = 2$, and the quartic case $\alf = 4$, which we have analyzed previously within RPA \cite{debbeler23}.

We have first computed the order parameter susceptibility and the fermion self-energy within RPA for generic values of $\alf$. The static susceptibility forms a sharp peak in momentum space at $\bQ$ for any $\alf > 2$, while for $\alf = 2$ the susceptibility becomes flat in a half-plane near $\bQ$ \cite{stern67}. The susceptibility (as a function of momentum and frequency) can be written in a scaling form with a scaling function that depends only on one variable. For $\alf = 2$ and $\alf = 4$ exact analytic expressions for the scaling functions can be obtained.
At the hot spots the (real) frequency dependence of the imaginary part of the fermion self-energy is linear and slightly asymmetric for any $\alf > 2$, while the real part exhibits a logarithmic divergence, which indicates a vanishing quasiparticle weight. The momentum dependence perpendicular to the Fermi surface also develops a logarithmic divergence for any $\alf > 2$. For $\alf = 4$ also the coefficient of the quartic momentum dependence diverges logarithmically, while for $\alf < 4$ only a finite correction to the prefactor of the leading $|k_t|^\alf$ term is obtained. For $\alf = 4$, these RPA results have already been derived in Ref.~\cite{debbeler23}, except for the exact analytic formula for the susceptibility.

A field-theoretic renormalization group analysis reveals that the logarithmic divergences in RPA are {\em not}\/ perturbative signatures of non-Fermi liquid behavior with anomalous power laws. Instead, only logarithmic corrections to Fermi liquid behavior are obtained. In particular, the quasiparticle weight and the Fermi velocity vanish logarithmically at the hot spots. Fermi liquids with a logarithmically vanishing quasiparticle weight are known as {\em marginal}\/ Fermi liquids \cite{varma89}.
The reason for this relatively mild breakdown of Fermi liquid behavior is the vanishing renormalized coupling between the electrons and the critical order parameter fluctuations, which is due to the suppression of spectral weight at the Fermi level.

The one-loop renormalization group analysis is controlled by a $1/N$ expansion, where $N$ is the number of fermion species. Higher loop orders correspond to higher orders in $1/N$. The one-loop fluctuation corrections to the fermion self-energy are numerically quite small for $N=2$ (that is, for electrons), as long as $\alf$ is not close to two. Feedback of the marginal Fermi liquid behavior of the self-energy on the susceptibility occurs only at the two-loop level. Except for $\alf \approx 2$, it is unlikely that this feedback will significantly affect the robust peak at $\bq = \bQ$ in the susceptibility. Hence, the $2k_F$ quantum critical point is not destroyed by fluctuations. This conclusion holds, in particular, for the physically realizable case $\alf = 4$.

For $\alf \to 2$, instead, the one-loop fluctuation corrections become large for any fixed finite $N$. Not only the RPA, but also the one-loop renormalization group becomes uncontrolled in this limit. Hence, for the parabolic case $\alf = 2$, we have not gained much further insight beyond the results of Refs.~\cite{sykora18} and \cite{halbinger19}.
The fluctuation induced flattening of the Fermi surface found for $\alf = 2$ in Refs.~\cite{sykora18,halbinger19} has been confirmed for $\alf > 2$, and may thus be a rather generic effect.


\begin{acknowledgments}
We are grateful to Elio K\"onig and J\"org Schmalian for valuable discussions.
\end{acknowledgments}


\begin{appendix}

\section{Evaluation of particle-hole bubble} \label{app:A}

The $k_r$ integral in Eq.~\eqref{Pi0def} can be easily carried out by using the residue theorem. Shifting the remaining integration variables as $k_t \to k_t + q_t/2$ and $k_0 \to k_0 + q_0/2$ to symmetrize the integrand, one obtains
%
\begin{align} \label{Pi0krint}
 & \Pi_0(\bq,iq_0) = \frac{i}{v}
 \int_{-\infty}^{\infty} \frac{dk_0}{2\pi} \int_{-\infty}^{\infty} \frac{dk_t}{2\pi} \\
 & \times \frac{\Theta(-k_0-q_0/2) - \Theta(k_0-q_0/2)}
 {2ik_0 - v q_r - b|k_t + q_t/2|^\alf - b|k_t - q_t/2|^\alf} \, .
\end{align}
%
Splitting the integral into contributions from positive and negative $k_0$ and shifting $k_0 \to k_0 \pm q_0/2$, the bubble can be written as
\begin{equation}
 \Pi_0(\bq,iq_0) = \Pi_0^+(\bq,iq_0) + \Pi_0^-(\bq,iq_0) \, ,
\end{equation}
where
%
\begin{align}
 & \Pi_0^+(\bq,iq_0) = - \frac{i}{v}
 \int_{-\infty}^{\infty} \frac{dk_t}{2\pi} 
 \int_0^{\infty} \frac{dk_0}{2\pi} \\
 & \times \frac{1}{2ik_0 + iq_0 - v q_r - b|k_t + q_t/2|^\alf - b|k_t - q_t/2|^\alf} \, , 
 \\[2mm]
 & \Pi_0^-(\bq,iq_0) = \frac{i}{v}
 \int_{-\infty}^{\infty} \frac{dk_t}{2\pi} \int_{-\infty}^0 \frac{dk_0}{2\pi} \\
 & \times \frac{1}{2ik_0 -iq_0 - v q_r - b|k_t + q_t/2|^\alf - b|k_t - q_t/2|^\alf} \, .
\end{align}
%
These expressions can be continued analytically to real frequencies by simply substituting $iq_0 \to \om + i0^+$. The justification for this step provided for the case $\alf = 4$ in Ref.~\cite{debbeler23} holds also for generic values of $\alf$.
Substituting $k_t = |q_t| \tk_t$ and $k_0 = b|q_t|^\alf \tk_0$, one obtains
%
\begin{align}
 & \Pi_0^{\pm}(\bq,\om) = \mp \frac{i}{v}|q_t|
 \int_{-\infty}^{\infty} \frac{d\tk_t}{2\pi} 
 \int_0^{\infty} \frac{d\tk_0}{2\pi} \\
 & \times \frac{1}{\pm 2i\tk_0 + \frac{\pm\om - v q_r}{b |q_t|^\alf} 
 - |\tk_t + \frac{1}{2}|^\alf - |\tk_t - \frac{1}{2}|^\alf} \, .
\end{align}
%
Subtracting $\Pi_0(\bQ,0)$ yields Eqs.~\eqref{dPi0}, with the scaling function
%
\begin{align}
 & I_\alf(x) = - 4i \int_{-\infty}^{\infty} \frac{d\tk_t}{2\pi} 
 \int_0^{\infty} \frac{d\tk_0}{2\pi} \\
 & \left[ \frac{1}{2i\tk_0 + x 
 - |\tk_t + \frac{1}{2}|^\alf - |\tk_t - \frac{1}{2}|^\alf}  
 - \frac{1}{2i\tk_0 - 2|\tk_t|^\alf} \right] .
\end{align}
%
Performing the frequency integral yields $I_\alf(x)$ in the form Eq.~\eqref{Ialf}.

For $\alf=2$ and $\alf=4$, the remaining integral over $\tk_t$ in Eq.~\eqref{Ialf} can be performed analytically. Here we present a derivation for the non-trivial case $\alf=4$.
We first factorize the numerator of the argument of the logarithm in Eq.~\eqref{Ialf} as
$(\tk_t + \frac{1}{2})^4 + (\tk_t - \frac{1}{2})^4 - (x + i0^+) =
 2(\tk_t^2 - \tk_+^2)(\tk_t^2 - \tk_-^2)$, where
\begin{equation}
 \tk_\pm = \sqrt{-\frac{3}{4} \pm \frac{1}{\sqrt{2}} \sqrt{1+x+i0^+}} \, .
\end{equation}
The logarithm in Eq.~\eqref{Ialf} can then be written in the form
$\ln\frac{\tk_t^2 - \tk_+^2}{\tk_t^2} + \ln\frac{\tk_t^2 - \tk_-^2}{\tk_t^2}$.
From $\int dk \, \ln k = k \ln k - k$ one easily derives
\begin{equation} \label{intlog}
 \int d\tk_t \, \ln\frac{\tk_t^2 - \tk_\pm^2}{\tk_t^2} =
 \tk_t \ln\frac{\tk_t^2 - \tk_\pm^2}{\tk_t^2} +
 \tk_\pm \ln\frac{\tk_t + \tk_\pm}{\tk_t - \tk_\pm} \, .
\end{equation}
The integrand is singular at $\tk_t = 0$ and, for real $\tk_\pm$, at $\tk_t = \tk_\pm$ and $\tk_t = - \tk_\pm$. However, the right hand side of Eq.~\eqref{intlog} is finite and continuous at $\tk_t = \tk_\pm$ and $\tk_t = - \tk_\pm$. Hence, the $\tk_t$ integral needs to be split only at $\tk_t = 0$, and we obtain
%
\begin{align}
 & \int_{-\infty}^{\infty} d\tk_t \, \ln\frac{\tk_t^2 - \tk_\pm^2}{\tk_t^2} = \\
 & \int_{-\infty}^{-0^+} d\tk_t \, \ln\frac{\tk_t^2 - \tk_\pm^2}{\tk_t^2} +
   \int_{0^+}^{\infty} d\tk_t \, \ln\frac{\tk_t^2 - \tk_\pm^2}{\tk_t^2} = \\
 & \tk_\pm \ln\frac{-0^+ + \tk_\pm}{-0^+ - \tk_\pm} -
   \tk_\pm \ln\frac{0^+ + \tk_\pm}{0^+ - \tk_\pm} = \mp 2\pi i \tk_\pm \, ,
\end{align}
%
where in the last step we have used $\Im\,\tk_+ > 0$ and $\Im\,\tk_- < 0$ (the imaginary part can be finite or infinitesimal).
Summing the contributions from $\tk_+$ and $\tk_-$, and using the relation
\begin{equation}
 i\tk_- - i\tk_+ = \sqrt{\frac{3}{2} + \sqrt{\frac{1}{4} - 2(x+i0^+)}} \, ,
 \nonumber
\end{equation}
one obtains the result Eq.~\eqref{I4x} for $I_4(x)$.


\section{Coefficients $A_s$, $B_s$, $C_s$, and $D_s$} \label{app:B}

\begin{figure}[tb]
\centering
\includegraphics[width=7cm]{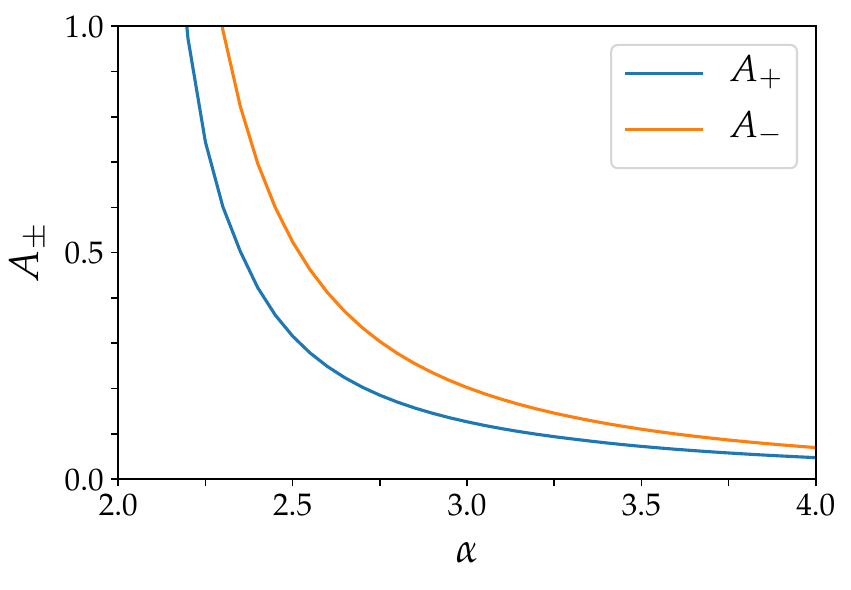}
\includegraphics[width=7cm]{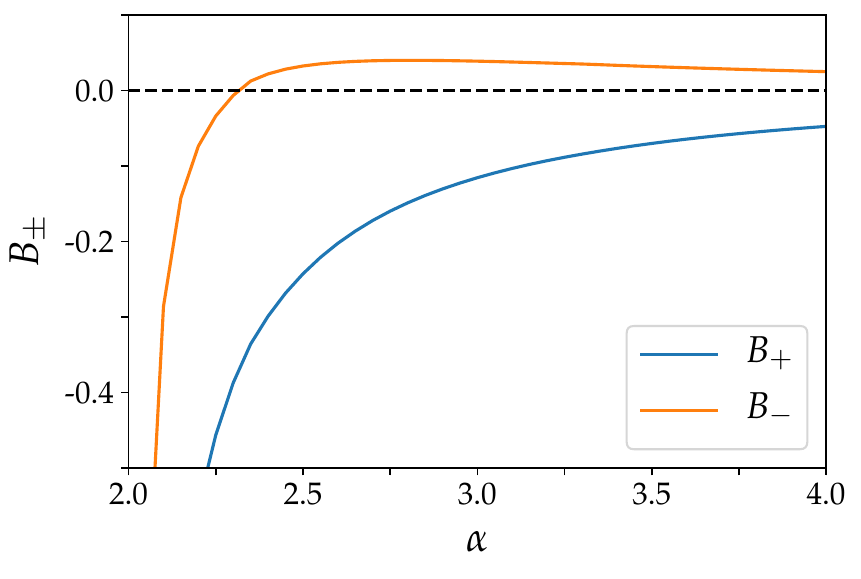}
\includegraphics[width=7cm]{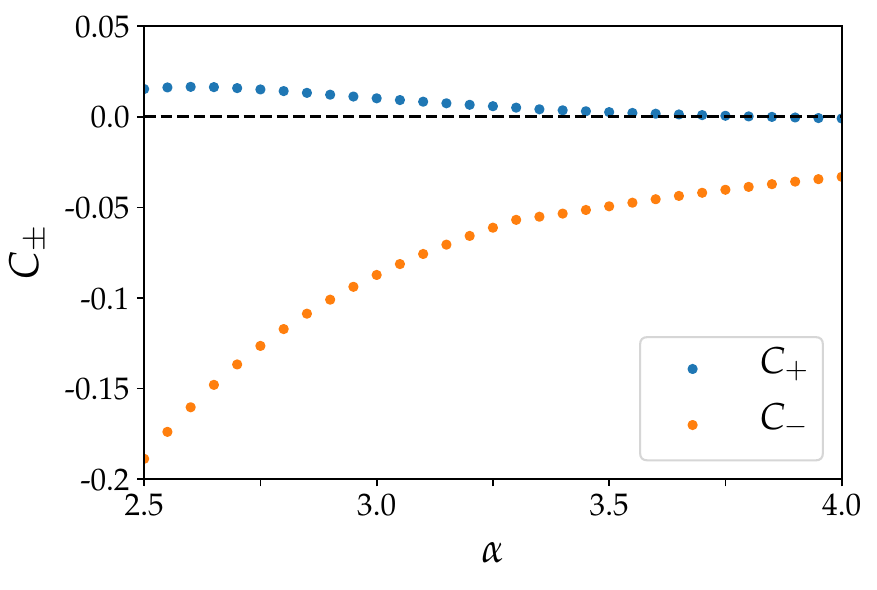}
\includegraphics[width=7cm]{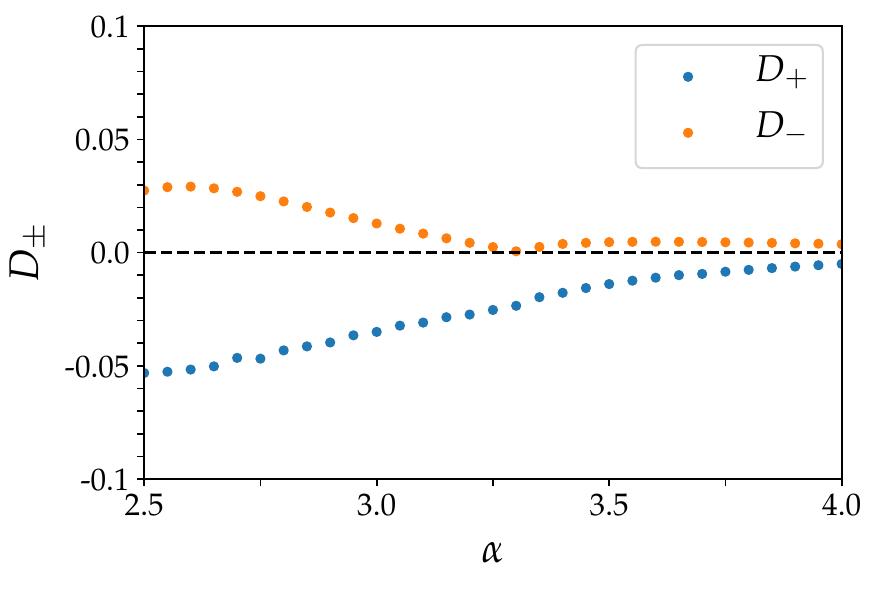}
\caption{Coefficients $A_\pm$, $B_\pm$, $C_\pm$, and $D_\pm$ as functions of $\alf$.}
\label{fig:coeff}
\end{figure}

In Fig.~\ref{fig:coeff} we show graphs of the expansion coefficients $A_s$, $B_s$, $C_s$, and $D_s$ defined in Sec.~III as functions of $\alf$.

\end{appendix}



\begin{thebibliography}{99}
\bibitem{fn_incomm} Commensurate wave vectors can be written as a linear
 combination of reciprocal lattice vectors with rational coefficients,
 while incommensurate wave vectors cannot. Obviously there are infinitly
 many possible rational coefficients, but the quantitative effects of
 commensurability decrease rapidly with the size of their denominators.
 In a one-loop calculation there is no difference between commensurate
 and incommensurate cases, as long as $2\bQ$ is not a reciprocal lattice
 vector.
\bibitem{footnote_At} In the plot of the scaling function $A_s^{(t)}(\tk_t)$ for $\alf = 4$ in Ref.~\cite{debbeler23}, a slightly different definition was used, corresponding to a stretched argument $\tk_t \to \sqrt[4]{2} \tk_t$.
\end{thebibliography}
\end{document}